\documentclass{emulateapj}
\usepackage{lscape}
\usepackage{epstopdf}
\usepackage{epsfig}
\usepackage{natbib}

\shorttitle{MASE}
\shortauthors{Bochanski et al.}
\slugcomment{PASP, accepted}

\begin{document}

\title{MASE:  A New Data--Reduction Pipeline for the Magellan Echellette Spectrograph}

\author{ John J. Bochanski\altaffilmark{1}, 
Joseph F. Hennawi\altaffilmark{2,3},
Robert A. Simcoe\altaffilmark{1}, 
J. Xavier Prochaska\altaffilmark{4},
Andrew A. West\altaffilmark{1},
Adam J. Burgasser\altaffilmark{1},
Scott M. Burles\altaffilmark{5},  
Rebecca A. Bernstein\altaffilmark{4}, 
Christopher L. Williams\altaffilmark{1},
Michael T. Murphy\altaffilmark{6}
}

\altaffiltext{1}{MIT Kavli Institute for Astrophysics \&
Space Research, 77 Massachusetts Ave, Building 37, Cambridge, MA
02139}
\altaffiltext{2}{Astronomy Department, University of California, 601 Campbell Hall, Berkeley, CA 94720-3411}
\altaffiltext{3}{Lawrence Berkeley National Laboratory, Berkeley, CA 94720}
\altaffiltext{4}{Department of Astronomy and Astrophysics, UCO/Lick Observatory, University of California, 1156 High Street, Santa Cruz, CA 95064}
\altaffiltext{5}{D. E. Shaw \& Co., L.P.,20400 Stevens Creek Boulevard,Suite 850, Cupertino, CA 95014}
\altaffiltext{6}{Centre for Astrophysics and Supercomputing, Swinburne University of Technology, Mail H39, PO Box 218, Victoria 3122, Australia}

\begin{abstract}
We introduce a data reduction package written in Interactive Data Language (IDL) for the Magellan Echellete Spectrograph (MAGE).  MAGE is a medium--resolution ($R \sim$ 4100), cross--dispersed, optical spectrograph, with coverage from $\sim$ 3000 -- 10000 \AA.   The MAGE Spectral Extractor (MASE) incorporates the entire image reduction and calibration process, including bias subtraction, flat fielding, wavelength calibration, sky subtraction, object extraction and flux calibration of point sources.   We include examples of the user interface and reduced spectra.  We show that the wavelength calibration is sufficient to achieve $\sim 5$ km s$^{-1}$ RMS accuracy and relative flux calibrations better than $10\%$.  A light-weight version of the full reduction pipeline has been included for real--time source extraction and signal--to--noise estimation at the telescope.
\end{abstract}


\section{Introduction}
Cross--dispersed spectrographs are attractive instruments for a variety of science applications, primarily for their broad wavelength coverage and high spectral resolution.   While these instruments maximize wavelength coverage and spectral resolution, the reduction and calibration of their data is complex.  The spatial and dispersion axes are typically not aligned with the CCD columns and rows; a given wavelength can span several columns on the chip.  Furthermore, the curvature and tilting of the orders complicates the tracing of order edges and the extraction of spectra.

Fortunately, many of these issues have been addressed by modern reduction techniques \citep[e.g.,][]{1982A&A...111..260M,1985A&A...143...13R,1986SPIE..627..707P,1988PASP..100.1572G, 1989PASP..101.1032M, 1990PASP..102..183M, 1994PASP..106..315H, 2002A&A...385.1095P, 2003PASP..115..688K}.  Pipelines exist and are regularly employed for extracting cross--dispersed optical \citep{2002A&A...385.1095P} and infrared \citep{2004PASP..116..362C} data.  Large, multi--fiber spectroscopic surveys, such as the Sloan Digital Sky Survey \citep[SDSS;][]{2000AJ....120.1579Y, 2002AJ....123..485S, 2002SPIE.4847..452S}, the Large Sky Area Multi--Object Fiber Spectroscopic Telescope \citep[LAMOST;][]{2008AcASn..49..327C} and the Radial Velocity Experiment \citep[RAVE;][]{2006AJ....132.1645S} also use pipeline reduction techniques that are directly applicable to the analysis of cross--dispersed data.

We have developed a new reduction package for the Magellan Echellete spectrograph \citep[MAGE;][]{2008SPIE.7014E.169M}.  The MAGE Spectral Extractor (MASE) is written in the Interactive Data Language (IDL) and is based on the SDSS spectro2d pipeline\footnote{Available at \url{http://spectro.princeton.edu/}}, the Magellan Inamori Kyocera Echelle \citep[MIKE;][]{2003SPIE.4841.1694B} IDL pipeline\footnote{Available at http://web.mit.edu/\~{}burles/www/MIKE/} (Bernstein, Burles \& Prochaska, in prep) and the XIDL astronomy package\footnote{Available at http://www.ucolick.org/\~{}xavier/IDL/index.html}.  The software employs a graphical user interface (GUI) built with IDL widgets to control the reduction process.  A faster, ``quicklook" version of the pipeline is packaged with MASE, for use at the telescope.   This version can be used for on--the--fly reductions and provides estimates of signal--to--noise ratios and spectral morphology.

In this paper, we describe the details of the MASE reduction pipeline and GUI.  In \S \ref{sec:mage}, we review the properties of the MAGE spectrograph.  The reduction pipeline is detailed in \S \ref{sec:mase}, with examples of the MASE GUI and reduced spectra.  Our conclusions are reported in \S \ref{sec:summary}.

\section{MAGE}\label{sec:mage}
\begin{figure*}[htbp]
\centering
\includegraphics[scale=0.20]{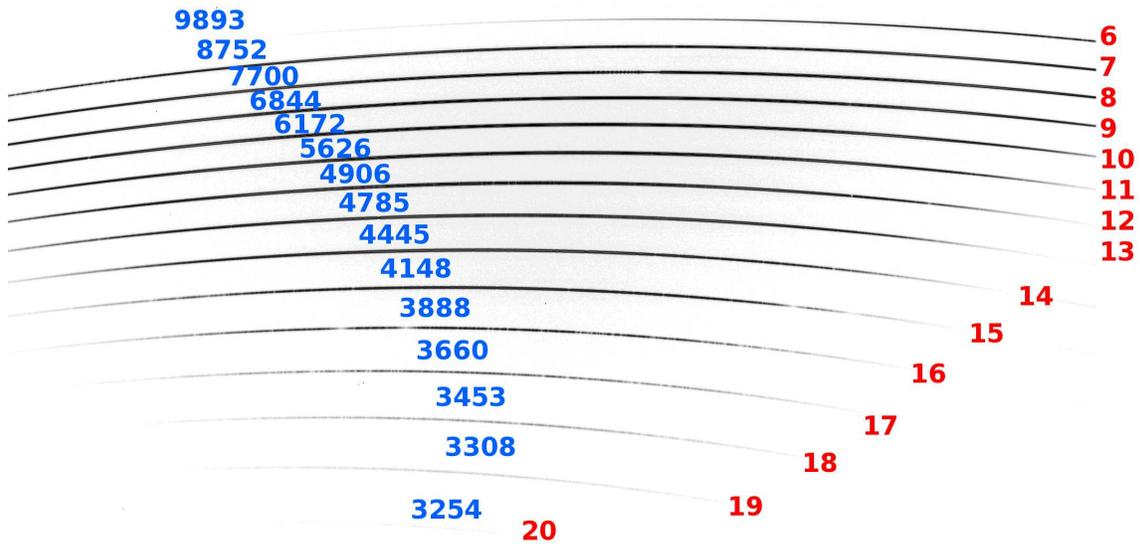}
\caption{An example MAGE spectrum of a flux standard.  The central wavelengths (blue) and order numbers (red) are listed near each order.  The position of the wavelength labels do not correspond to their actual location on the chip.  Note that the sensitivity of MAGE falls off sharply in orders 6 and 20.}
\label{fig:example}
\end{figure*}

The MAGE spectrograph is described in detail by \cite{2008SPIE.7014E.169M}, but we briefly summarize details critical to the reduction and calibration process here.  MAGE is a medium--resolution ($R \simeq 4100$, 22 km s$^{-1}$ pixel$^{-1}$) optical spectrograph currently mounted on the central folded port of the Clay telescope at the Las Campanas Observatory in Chile.   Similar to the Echellette Spectrograph and Imager \citep[ESI;][]{2002PASP..114..851S} on the Keck II telescope,  MAGE employs a grating in combination with two prisms to provide cross dispersion.  The prisms are aligned to minimize anamorphic magnification.   The instrument's wavelength coverage spans 3000 - 10500 \AA\ over 15 orders ($6 < m < 20$), continuous except for a 15 \AA\ gap near 9500 \AA\ between orders six and seven.  Five non--overlapping slits are available, with widths from $5.0^{\prime\prime}$ to $0.5^{\prime\prime}$,  corresponding to resolutions of $R \sim 1000$ ($\sim$ 120 km s$^{-1}$) to $R \sim 8000$ ($\sim$ 12 km s$^{-1}$).  Each slit is $10^{\prime\prime}$ long.  The slit is imaged with a $f/$1.4 Schmidt vacuum camera equipped with an E2V back--illuminated CCD.  The chip is composed of 2048 $\times$ 1024 13.5 $\mu$m pixels.  Each pixel subtends 0.33$^{\prime\prime}$ on the sky.  Dispersion on the chip ranges from 0.3 \AA$~$pixel$^{-1}$ (blue) to 0.6 \AA$~$pixel$^{-1}$ (red).  A typical MAGE science observation is shown in Figure \ref{fig:example}.  

Two calibration sources are incorporated into MAGE.  A Thorium-Argon cathode tube is used for wavelength calibration, producing arc lines over the entire spectral window.   ThAr arcs should be acquired after each science target to establish an accurate wavelength solution.    A Xenon--flash bulb is used for flat fielding in the bluer orders ($m \geq 14$), but this lamp cannot be used to flat field the redder orders due to the presence of strong emission lines.  We recommend the following calibration images to construct pixel and illumination flats. Pixel flats are constructed from three sets of observations.  First, for the bluest orders (18 - 20), in focus twilight flats should be acquired with the 5.0$^{\prime\prime}$ slit.  This wide slit diffuses the Fraunhoffer lines in the sky spectrum. For the green orders (14 - 17), the Xe-flash lamp observed with the science slit provides suitable illumination for flat fielding.  Finally, in the red orders (6 - 13), the bright quartz lamp and dome screen observed with the science slit provide suitable illumination of the optical path.  For each set of orders, at least 10 focused flat--field images are recommended to obtain a reliable median.  Illumination flats should also be acquired during twilight with the science slit. Since the prisms and gratings within MAGE do not move, the instrument is very stable and only one flat field is typically needed per setup per night.  If slits are changed during the night, a new set of flats should be acquired after the change.  Flux calibration is obtained by observing spectrophotometric standards \citep[e.g., ][]{1990AJ.....99.1621O, 1992PASP..104..533H, 1994PASP..106..566H, 1988ApJ...328..315M}.  At least one standard per setup should be observed during the run.  Note that observations of flux calibrators are also required for object tracing. A table describing the necessary MAGE calibrators is listed below (Table \ref{table:cals}).
 
\begin{center}
\begin{deluxetable*}{lrrrr}

 \tablecaption{Suggested Calibrations}
 \tablehead{
 \colhead{Calibration Source} &
 \colhead{Slit} &
 \colhead{Applicable Orders} &
 \colhead{Exposure Time\tablenotemark{a}} &
 \colhead{Num. of Exp.}
 }
\startdata
Quartz Lamp & Science Slit & 6-13 & 40s & $\gtrsim 10$\\
Xe-Flash Lamp & Science Slit & 14-17  & 35s & $\gtrsim 10$ \\
Twilight & 5$^{\prime\prime}$ & 18-20 & $\sim$ 60s \tablenotemark{b} & $\gtrsim 10$ \\
Twilight\tablenotemark{c} & Science Slit & 6-20  & $\sim$ 60s & $\gtrsim 1$\\
Flux Calibrator & Science Slit & 6-20 & Varies & $\gtrsim 1$ \\
Thorium Argon & Science Slit & 6-20  & 3s & Varies \\
\enddata
\label{table:cals}
\tablenotetext{a}{Exposure times are given for the 0.7$^{\prime\prime}$, and should be scaled down for larger slits.}
\tablenotetext{b}{Twilight exposure times should be adjusted for changing sky conditions.}
\tablenotetext{c}{These exposures are used for illumination flats.}
\end{deluxetable*}
\end{center}

\section{Data Reduction}\label{sec:mase}

\subsection{User Interface}\label{sec:gui}
\begin{figure*}
\includegraphics[scale=0.55]{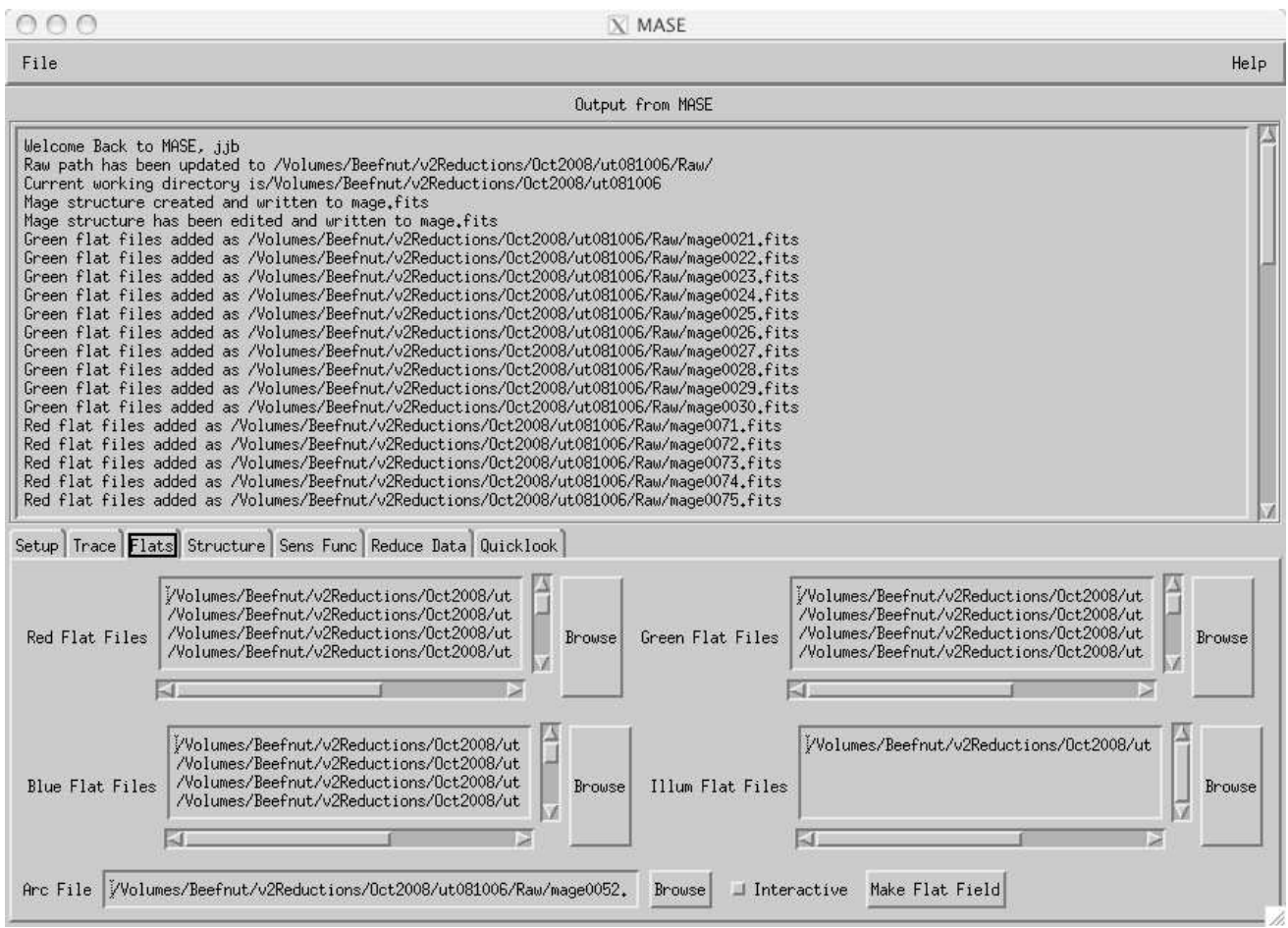}
\caption{The MASE GUI.  It is composed of the message window and tabs.  The message window, located in the upper part of the GUI, provides feedback and confirmation of user actions.  Underneath the message window, the tabs separate the major steps in the calibration process and allow the user to select which files are calibrated and reduced.  The Flats tab is shown in this example.}
\label{fig:mase_gui}
\end{figure*}

The MASE GUI is based on IDL widgets.   It is composed of two parts, the message window, which provides a log of the user's file selections and procedures executed, and the tabs, which allow for easy navigation through each step of the reduction process.  An example of the MASE GUI is shown in Figure \ref {fig:mase_gui}.  Within the tabs, the user specifies the current working directory, constructs calibration images, and executes the main reduction software.  Most common options, such as choosing which target to reduce, are available within the GUI interface.  Other options are accessible only from the underlying IDL code, which can be executed from the command line.  The default choices are optimized for point source extraction.   The MASE GUI and associated IDL programs are publicly available\footnote{The MASE website is located at \url{http://web.mit.edu/jjb/www/mase.html}}.

\begin{figure*}
\centering
\includegraphics[scale=0.68]{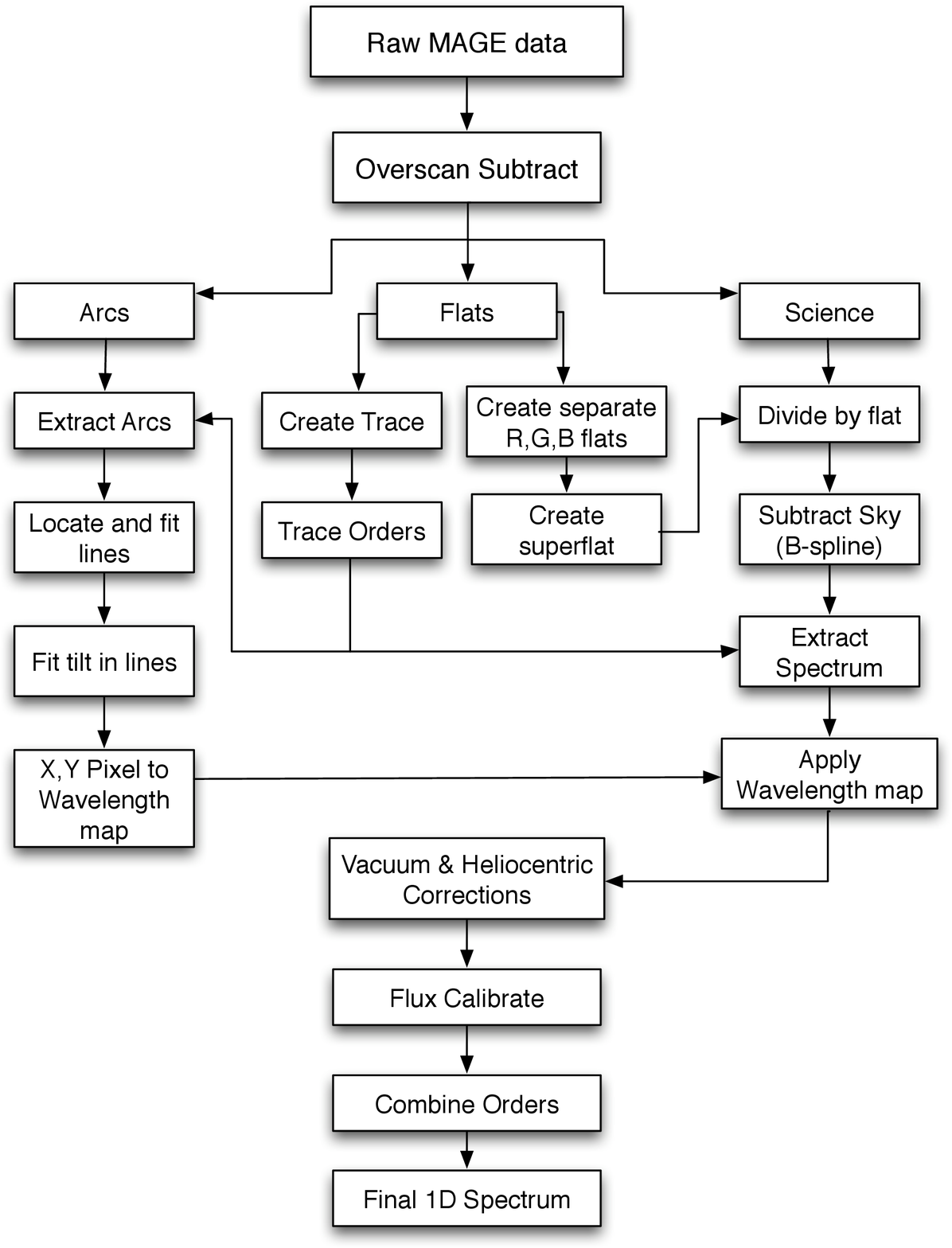}
\caption{An illustrative flowchart of the MASE reduction pipeline.  Each major step in the pipeline discussed in the text is shown.}
\label{fig:flowchart}
\end{figure*}

\subsection{Organization and Structure}
For each night of observations, MASE generates an IDL data structure to organize the reduction.  When the structure is first created, the program attempts to guess each type of exposure from header information and derived quantities. Each science target is assigned an object ID number and associated with the nearest temporal ThAr arc.  The structure is written out to a FITS file, and can be inspected and edited with utilities included within MASE.  

The user selects the directory containing the raw MAGE frames to begin the reduction process.  This selection is echoed in the MASE message window.  This data directory should be placed within a working directory (i.e. for each night's observing).  A series of subfolders are created in directories parallel to the raw directory to house files generated during the reduction process.

\subsection{Calibration:  Bias--subtraction}
The data reduction pipeline, outlined in Figure \ref{fig:flowchart}, begins with bias subtraction of every raw frame.  The median bias level is measured from a 128 $\times$ 128 pixel overscan region on the MAGE detector and subtracted from the entire image.  Each image is then trimmed to 2048 $\times$ 1024 pixels and pixel values are converted from ADUs to electrons, using a previous empirical measurement of the gain, which is recorded in the image header.  The predetermined read noise and electron counts (shot noise) are used to construct an inverse variance image, which is a first estimate of the uncertainty in the counts per pixel position caused by photon counting noise and detector read noise.

\subsection{Calibration: Flat fielding and Order Definition}
\begin{figure*}[htbp]
\centering
\includegraphics[scale=0.5]{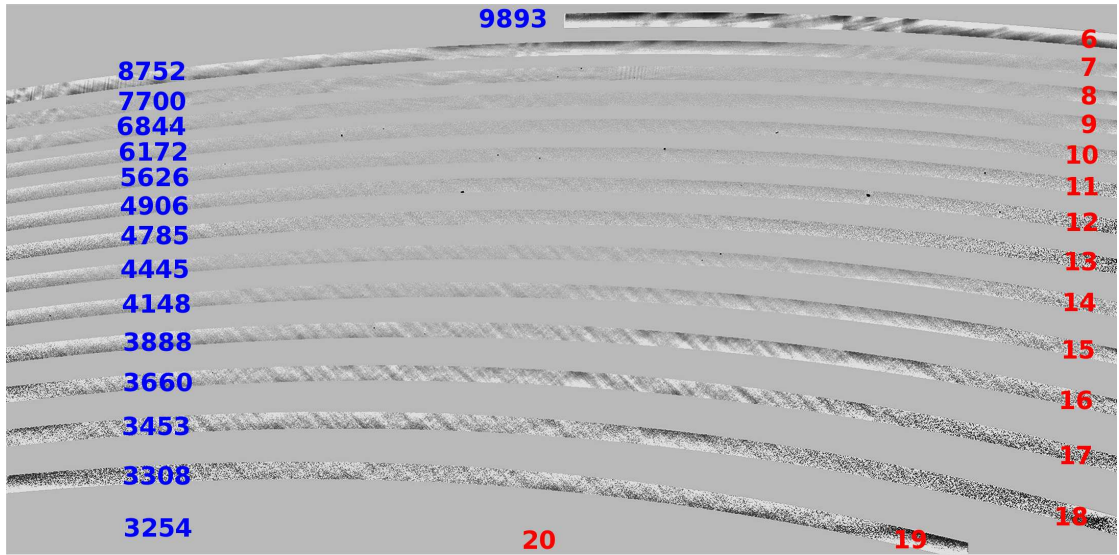}
\caption{An example ``superflat" combining separate red, green and blue flats.  The image has been stretched to enhance the contrast.  As in Figure \ref{fig:example} the order numbers are given in red and the central wavelengths of each order are listed in blue.  There is significant fringing in the reddest orders, and note that order 20 has all pixels set to unity.}
\label{fig:flatfield}
\end{figure*}

A flat--field image is constructed for each illumination source described in \S \ref{sec:mage}.  First, a median box filter is applied to each flat, smoothing the image over 30 pixel scales and removing large scale variations.  Each unsmoothed flat field image is then normalized by its median filtered counterpart.  The normalized flat--field images are then coadded and further filtered, where outlying pixels with deviations larger than two standard deviations from the mean are not included in the final coadd.   Next, the separate red, green and blue flats are merged into a single ``super flat".  The red flat is used for orders 6--13, the green flat spans orders 14-17, and the blue flat covers orders 18 through 20.  For typical cross--dispersed flat fields, the order edges can exhibit sharp discontinuities.  Each order was trimmed by four pixels on its upper and lower bound to minimize this effect.  Furthermore, inter--order pixels were set to unity.  An example of a super flat is shown in Figure \ref{fig:flatfield}.  The imaged is stretched to greatly enhance the contrast of the image and highlight features within the flat--field image.  In orders 6 and 7, fringing at the level of $\sim 10\%$ is clearly visible.  Orders 8 through 14 are very flat, showing no almost no variations, except for a few small ``holes''  (i.e., such as the one visible in order 12).  Orders 16 and 17 display some variations due to small fluctuations in the anti--reflective coating on the CCD.  Finally, order 20 is set to unity, as obtaining the required number of counts ($\sim$ 2-3 $\times$ 10$^3$)  at these wavelengths ($<$ 3300\AA) for flat fielding is severely hampered by atmospheric extinction.  The illumination correction is computed by fitting the variation in the spatial direction of a smoothed twilight flat field along the $x$ coordinate of each order.
 
A critical step in reducing cross--dispersed data is the identification and measurement of order edges.  Trace flats are constructed by summing red, green and blue flat--field images for each slit used. The trace image is convolved with a sawtooth filter, enhancing the order edges.  A fourth order Legendre polynomial is then fit to the upper and lower edges of each order as a function of column ($x$ pixel position).  The fit coefficients define the order edges, centers and widths on the chip.  The order information is output to a FITS file, and the fit is displayed for visual confirmation by the user.

\begin{center}
\begin{deluxetable}{lrrr}
\tablewidth{2.5in}
 \tablecaption{Wavelength ranges by order}
 \tablehead{
 \colhead{Order} &
 \colhead{$\lambda_{min}$} &
 \colhead{$\lambda_{central}$} &
 \colhead{$\lambda_{max}$} 
 }
 \startdata
 6 & 9500 & 9893 & 10285\\
 7 & 8128 & 8752 & 9475\\
 8 & 7112 & 7700 & 8288\\
 9 & 6320 & 6844 & 7368\\
10 & 5713 & 6172 & 6630\\
11 & 5225 & 5626 & 6027\\
12 & 4286 & 4906 & 5525\\
13 & 4469 & 4785 & 5100\\
14 & 4170 & 4445 & 4720\\
15 & 3905 & 4148 & 4390\\
16 & 3675 & 3888 & 4100\\
17 & 3467 & 3660 & 3853\\
18 & 3287 & 3453 & 3620\\
19 & 3187 & 3308 & 3430\\
20 & 3197 & 3254 & 3311\\
\enddata
 \label{table:orders}
\end{deluxetable}
\end{center}

\begin{figure*}[htbp]
\centering
\includegraphics[scale=0.20]{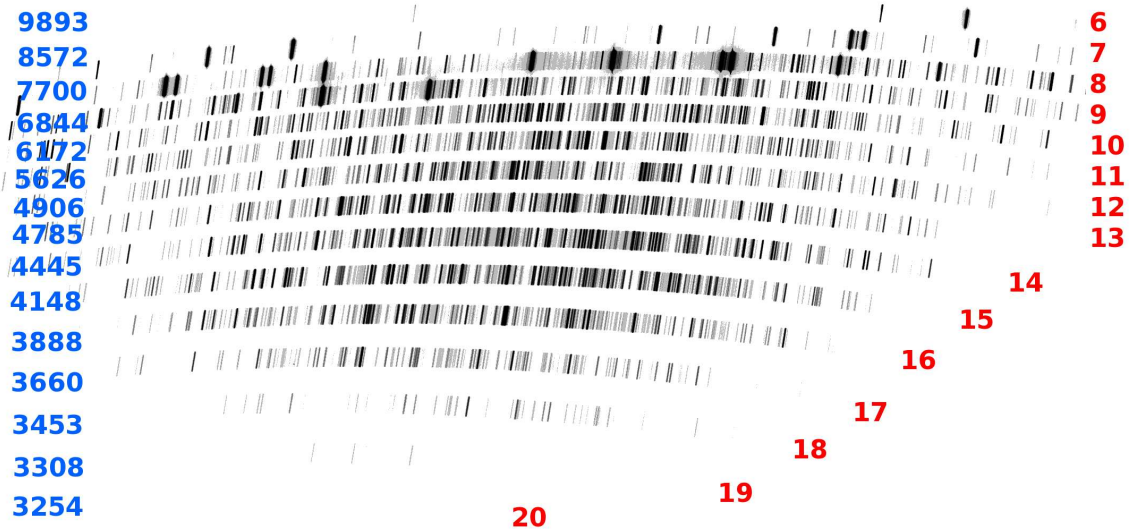}
\caption{Shown is an example ThAr arc spectrum.   Note the dearth of emission lines in the reddest and bluest orders.}
\label{fig:arcs_example}
\end{figure*}

\subsection{Calibration: Wavelength Solutions}
A wavelength solution is computed from each ThAr arc image.  An example arc frame is shown in Figure \ref{fig:arcs_example}.   Because the dispersion of light is not aligned along rows and columns, each pixel corresponds to a specific wavelength.  The wavelength solution is derived in two parts.  First, the pixels at the center of each order are mapped to wavelength.   A boxcar extraction down the center of each order is used to extract a 1D arc spectrum in air wavelengths, which is compared to a ThAr line atlas \citep{2007MNRAS.378..221M}.   We note that many currently available line lists are appropriate for higher resolution echelle spectrographs (i.e., MIKE).  To construct a clean line list for MAGE, we removed line blends by eye, using the line atlas of \cite{2007MNRAS.378..221M} smoothed to MAGE resolutions as a guide.
An archived solution was produced by manually identifying emission lines in every order.  A Fourier transform is used to calculate a shift between the extracted arc spectrum and the archived wavelength solution.  These shifts are incorporated within an initial guess at the wavelength solution.  
 The line centroids of the extracted arc spectrum are identified and a low--order Chebyshev polynomial is fit to $\lambda$ vs. column ($x$ pixels) using least squares.   This result is used to identify slightly weaker lines excluded in the original fit, and the polynomial is re-fit.   Next, a two-dimensional solution is computed, fitting the product of the wavelength and echelle order as a function of CCD row and order number.  The 2D solution allows for interpolation through orders with a small amount of clean ThAr lines, such as orders 6 and 20.  It also constrains the wavelength solution to be the same in regions of overlap, where the 1D solutions are totally independent.  The 2D solution is not complete, as it only maps the center of each order to wavelength.  The tilt of the emission lines, which varies with position on the chip, must be incorporated into the final wavelength calibration.   In each order, strong emission lines are fit with a straight line, and the resulting slope is used to determine the tilt of the arcs as a function of position.  This information is combined with the center-order wavelength solution, and each science pixel is mapped to wavelength.  A 2D FITS image of the pixel-wavelength map is written in the Arcs directory.  

\begin{figure*}[htbp]
\centering
\includegraphics[scale=0.42, angle=90]{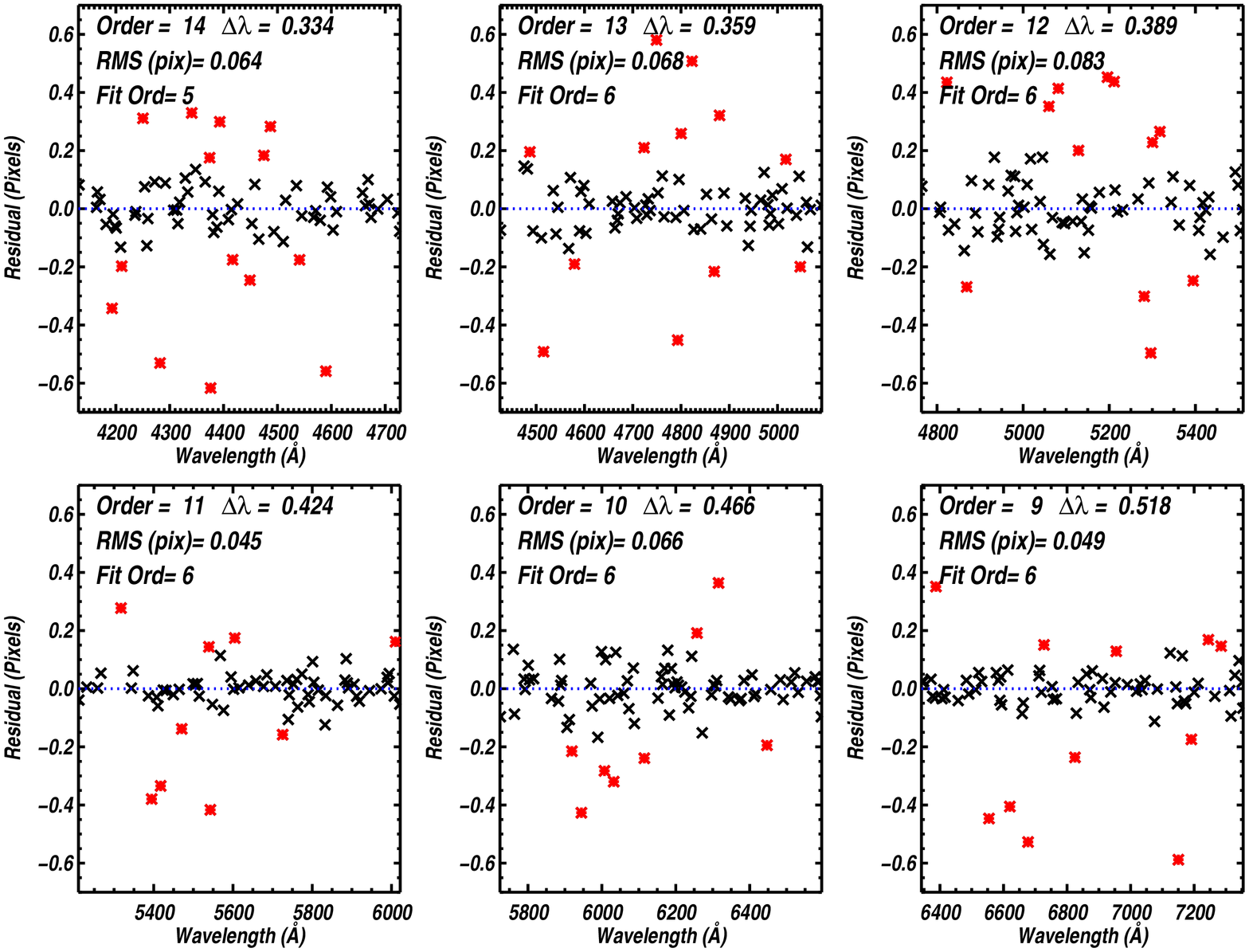}
\caption{An example of the QA plots produced during 1D wavelength calibration.  Plotted in each panel are the residuals (in pixels) as a function of wavelength, for each order.  The black crosses correspond to ThAr emission lines that were used in the final fit, and red asterisks are lines that were rejected.  The order number, RMS (in pixels),  dispersion (\AA$~$pixel$^{-1}$) and polynomial order is reported for each panel.  Note that the typical RMS is less than 0.1 pixels and usually $\sim$ 0.05 pixels.}
\label{fig:arcs_qa}
\end{figure*}

\begin{figure*}[htbp]
\centering
\includegraphics[scale=0.35, angle=90]{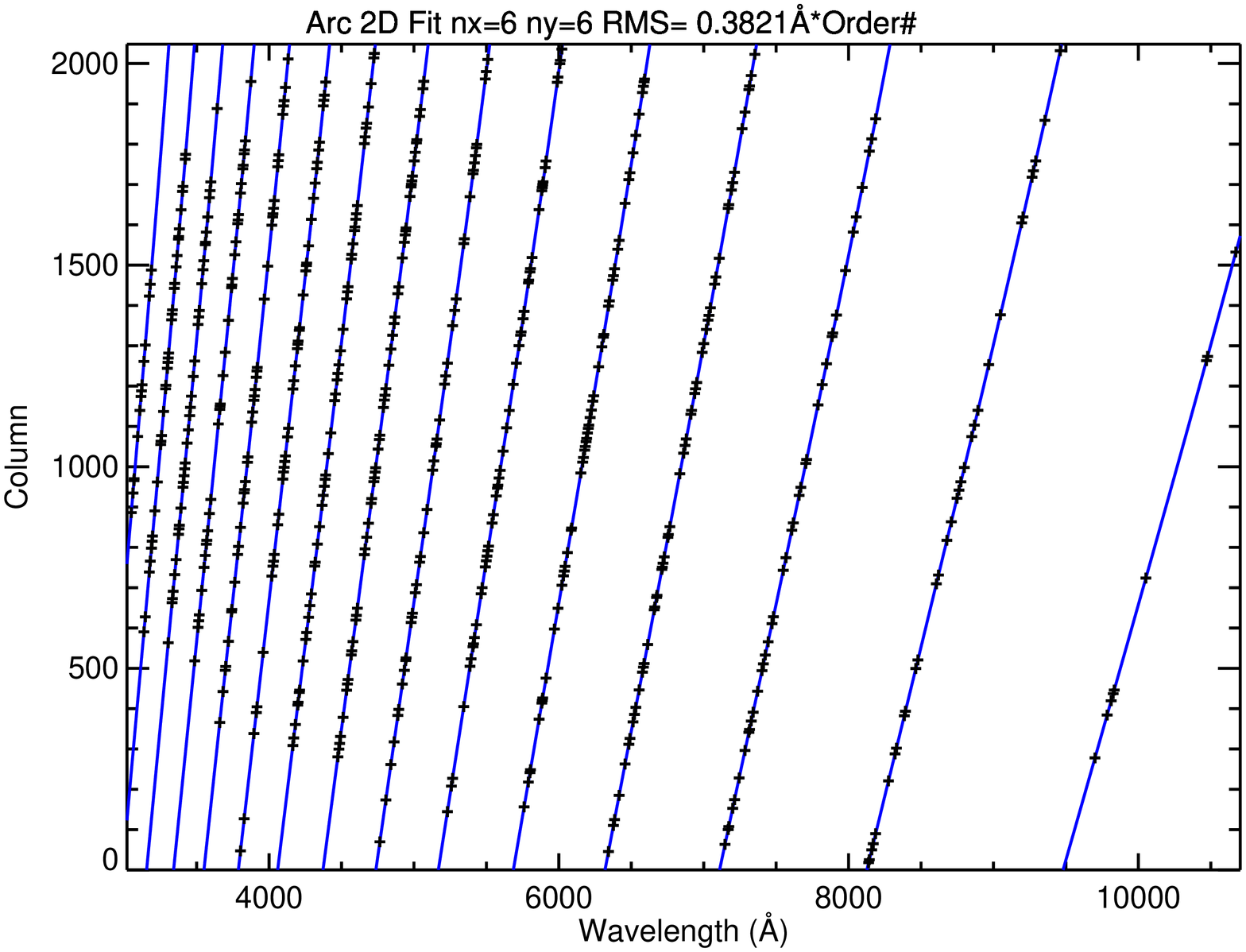}
\caption{An example of the QA plots produced during 2D wavelength calibration. The blue line is the 2D Chebyshev polynomial fit, and the crosses are the individual line centers, shown as a function of column ($x$ pixel position) and wavelength.  This step improves the accuracy in orders with few ThAr lines, such as order 6 (the rightmost order).}
\label{fig:twod_arcs_qa}
\end{figure*}

Multiple quality assurance (QA) plots are produced during the wavelength calibration process.   This allows the user to inspect both the one dimensional wavelength solution for each order and the two dimensional fits.  The root mean square (RMS) deviation in pixels is recorded for each order, and lines that are rejected from the final fit are flagged.   Example QA plots are shown in Figures \ref{fig:arcs_qa} and \ref{fig:twod_arcs_qa}.  Typical RMS values for the one dimensional fits are $\sim$ 0.05, and ranges up to 0.2 pixels for orders with very few ThAr lines (6 \& 20).   This corresponds to roughly 0.02 - 0.07 \AA\ and sets the velocity precision of the MAGE spectrograph. This precision is approximately 5 km s$^{-1}$, as verified by cross--correlating observations of known TW Hydrae Association members using the 0.7$^{\prime\prime}$ slit with known radial velocity standards \citep{2007AJ....133..531B}.

MAGE is an exceptionally stable instrument, especially over the course of a night of observations.  To test of the wavelength solution stability, we extracted 13 sky spectra from a given night of observations.  No heliocentric corrections were applied, and the telluric emission lines were cross-corellated against each other.  The relative shifts over course of the night were typically less than 1 km s$^{-1}$, with a mean of $>$ 0.3  km s$^{-1}$.  This shift corresponds to $>$ 0.01 pixel, indicating MAGE is a very stable instrument.  Thus, the observer can confidently apply an arc taken before or after an observation, but can avoid the overhead cost of two arcs. 

\subsection{Extraction:  Sky--Subtraction, Object Finding, Aperture Tracing and Extraction}
Using the calibration images described above, each science target is overscan subtracted and flat fielded.   Yet, night sky (telluric) emission lines and background remain.  Most traditional sky--subtraction algorithms model telluric features after rectifying the observed spectra onto a uniform wavelength grid.  Unfortunately, rebinning the spectrum correlates the noise in the image, combining pixels that were previously independent.  However, recent data reduction pipelines \citep[e.g.,][]{2007NJPh....9..443B} have introduced basis splines \citep[B--Splines;][]{2003PASP..115..688K}, or piecewise polynomial solutions defined by an order and breakpoints to produce a model of the sky emission and remove it from the science image.  At each breakpoint, the B--spline is continuous through $n-2$ derivatives, where $n$ is the order.   Typically, 4th order B--splines are employed for sky subtraction, which use cubic piecewise functions.  B--splines are flexible enough to fit the night sky emission features as a function of CCD pixels, rather than wavelength.  This greatly simplifies the error analysis and propagation, and results in near Poisson-limit sky subtraction.  Additional discrepant features, such as cosmic rays or bad pixels are easily identified and removed from the model sky image. 

To start the extraction process, the object is crudely traced and masked.  Each order is collapsed along wavelength, and the spatial center of the profile is located with a median. The region along this crude trace is masked out, and a sky model is constructed from the background.  For each science target, a 4th order B--spline is fit as function of pixel position along a trace offset from the order center.  Using the tilt mapped during the wavelength calibration, the initial sky model is propagated across the entire spatial extent of the slit.  This initial sky model is used as a best guess for the final sky, which is later output as a FITS image in the Final directory.
 
Next, an object finding code is run to refine the crude object trace above. As described above, the code uses the trace of a previously reduced flux standard as a best guess for the object's position.  Since the flux standard is bright, locating and extracting its spectrum is straightforward and does not require a precise initial guess of its position on the chip.  First, the trace of each object is compared to the standard star trace in each order, and the distance between the science target and flux standard in each order is computed.   The traces are grouped by a friends--of--friends algorithm.  For orders in which a trace was not detected (i.e., a Lyman--limit system or T dwarf)  the median distance from the standard star trace (for all detected orders) is used.  Note that this does not account for any shift due to atmospheric refraction between the standard and science object. Once the trace center has been determined, the flux--weighted mean position as a function of CCD row is measured, and a low--order Legendre polynomial is fit to trace on the chip.  A structure, similar to the one describing the order edges, is generated to describe the location of the object along the chip.

\begin{figure*}[htbp]
\centering
\includegraphics[scale=0.4, angle=90]{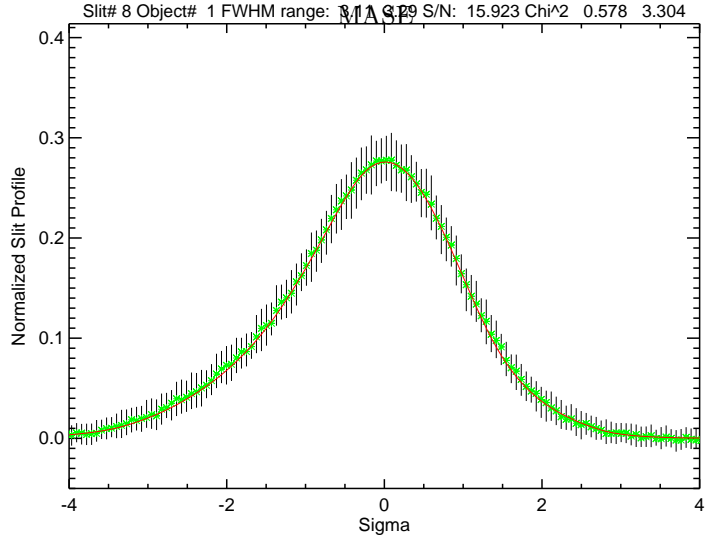}
\caption{An illustrative slit profile produced during extraction.  The normalized median counts are plotted as green crosses.  The error bars represent the $20^{\rm th}$ and $80^{\rm th}$ percentiles.  The fitted model is shown in red.  The profile is shown as a function of sigma standard deviations where sigma is defined in terms of the mean full width at half maximum ($\sigma = $ FWHM/2.35).   Note that the FWHM can vary along the trace.}
\label{fig:slit_profile}
\end{figure*}

After finding the object, the dispersion axis is collapsed and the counts are plotted against fractional slit height.  An example of the collapsed slit profile is shown in Figure \ref{fig:slit_profile}.   As the imaged slit width in pixels varies between orders, identifying the object in relative slit coordinates simplifies the extraction code.  Once the object trace is located on the chip, a two-step process is used to measure its flux.  First, a simple boxcar extraction is used to quickly estimate the signal--to--noise ratio (SNR) in each order.  The orders are then ranked by SNR, and an optimal extraction is used to extract the flux from the point source spectrum.  Optimal extraction was introduced by \cite{1986PASP...98..609H} and \cite{1986PASP...98.1220R} and updated for use in cross dispersed data by \cite{1990PASP..102..183M} and \cite{1989PASP..101.1032M}.  Briefly, optimal extraction weights each pixel by the fraction of the total flux impinging on that pixel.  These weights are calculated from the slit profile, which is measured along the collapsed order (seen in Figure \ref{fig:slit_profile}).   The profile is fit with a B-spline for high SNR observations, and for low SNR orders (with SNR $<$ 2.49), the profile parameters determined in the higher SNR orders are assumed.  If the median SNR of the spectrum falls below the limit, a Gaussian is assumed.  Note that optimal extraction is ideal for spectroscopy of unresolved point sources, and will recover the proper profile for extended objects as well, provided their slit profile shape does not vary strongly with wavelength.

Extraction takes place after the object profile and optimal extraction weights have been defined.  B--splines are employed to simultaneously model both the sky and object spectra.  This routine improves the model sky estimate described above.  After the first B-spline fit, pixels with a deviation greater than 3.5 sigma from the model are identified.  This mask is designed to remove bad pixels and cosmic ray hits.  The fit is repeated, with outliers masked out.   The two dimensional sky and object images are recorded, and the spectrum of the object is written out to a file.  The 2D images can be inpected with a separate IDL routine (mage\_look), which overlays image masks, order edges and the object trace on the 2D image.  The wavelength solution determined above is adopted during the extraction, after applying a helio-centric correction and converting to vacuum coordinates.

\subsection{Post Extraction:  Combining Orders and Flux Calibration}
After extraction, each order's spectrum is recorded in counts vs. wavelength. The spectra are then rebinned onto a common logarithmic wavelength grid.   Prior to this step, each pixel on the chip has been treated independently, which avoids correlated errors.  However, a common wavelength grid is necessary for co-adding multiple exposures, which occurs after rebinning.  Different exposures of a common science target are co-added by default, but the user has the option of recording individual spectra, useful for monitoring a given target (i.e., for radial velocity variations). 

\begin{figure*}[htbp]
\centering
\includegraphics[scale=0.75]{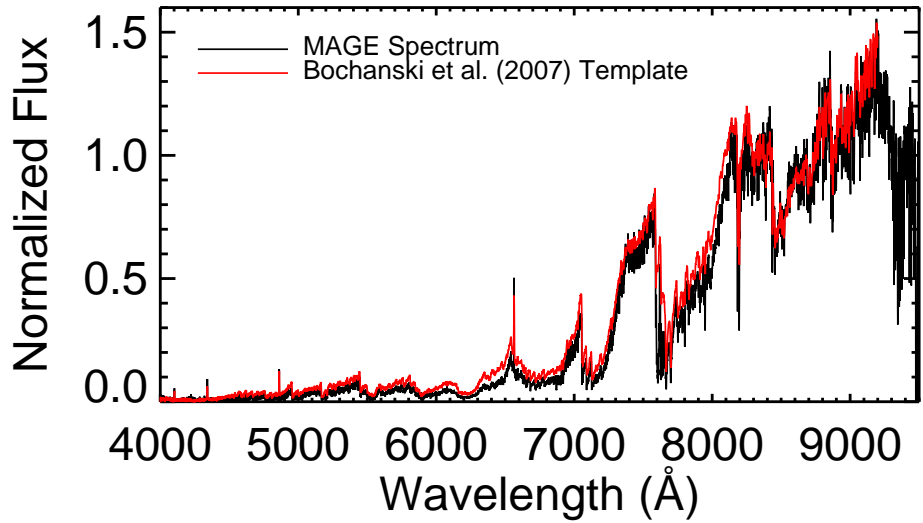}
\caption{Comparison between a reduced MAGE spectrum of an M7 dwarf (black line) and the corresponding SDSS template spectra \citep[red line;][]{2007AJ....133..531B}.  Note the fairly good agreement in the flux calibration between the two objects over the wavelength range shown, despite residual telluric absorption in the MAGE spectrum.  Future versions of MASE will include a telluric correction.}
\label{fig:comp}
\end{figure*}

After co-addition, a flux calibration is applied to the data.  The user can specify an archival solution or determine a new sensitivity function.  The standard star spectrum is extracted as described above, then compared to spectrophotometric atlases incorporated with MASE \citep[][]{1990AJ.....99.1621O, 1992PASP..104..533H, 1994PASP..106..566H, 1988ApJ...328..315M}.  A list of the standards included can be found on the MASE website.  The user can interactively mask out bad regions or strong absorption lines in the flux calibrator's spectrum.   A sensitivity function is determined (counts / flux as a function of wavelength) and applied to the science spectrum.   Typical relative fluxes are good to $\sim$ 10\%, as shown in Figure \ref{fig:comp}.  A FITS file containing the spectrum for each order is output to the Object directory.

\begin{figure*}[htbp]
\centering
\includegraphics[scale=0.75]{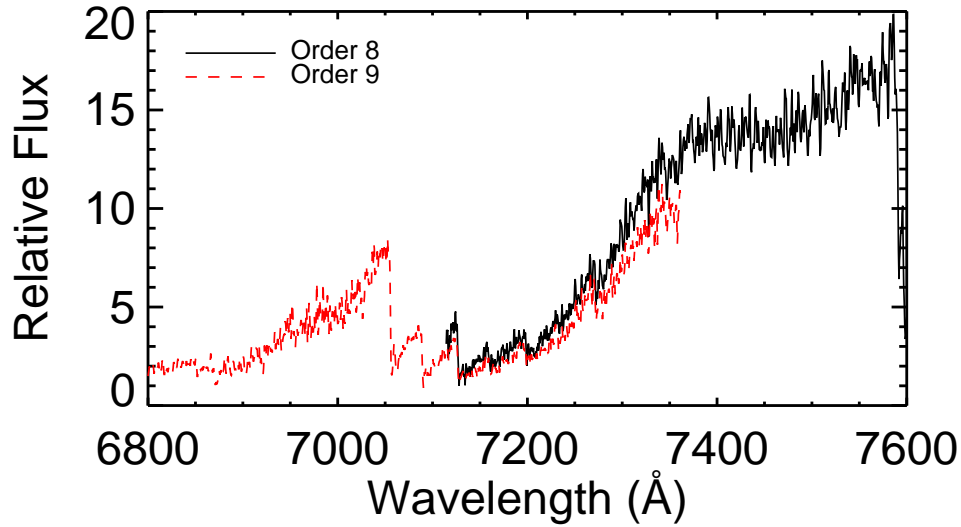}
\caption{The relative flux of an M dwarf observed with MAGE between two adjacent orders (8 \& 9, black solid line and red dashed line, respectively).  Note that the agreement between orders is very good, typically to within 5\%.}
\label{fig:overlap}
\end{figure*}

Finally, each co-added spectrum is combined to a one-dimensional flux vs. wavelength spectrum.  At the overlapping edges of sequential orders, the average is computed and recorded for that wavelength bin.   Agreement between the orders is usually good to $\lesssim 5\%$, as shown in Figure \ref{fig:overlap}. The 1D flux and error spectra are output to the FSpec directory.

\subsection{Quicklook Tool}
A  ``slimmed--down" version of the full pipeline is included with MASE for use at the telescope.  This quicklook tool (QT) can process a raw CCD image in a few minutes and allow the observer to quickly estimate signal--to--noise and inspect spectral morphology.  It employs archival order edge traces, wavelength solutions and sensitivity functions.  The QT uses boxcar extraction instead of optimal extraction, resulting in major improvements in speed.  However, emission lines are commonly mis-identified as night sky lines and subtracted from the spectrum.  The user should always use the full reduction package for scientific analysis.  The QT is accessed by clicking on the ``Quicklook" tab in the MASE GUI.  The only user input is the filename of the science exposure.  If the user does not have MASE installed on their own computer, the QT is also installed locally at Las Campanas.

\section{Summary}\label{sec:summary}
We have developed a new data reduction package for the MAGE spectrograph.  The MASE GUI and pipeline is written in IDL and incorporates modern sky--subtraction and optimal extraction techniques.   The program is available for download online.  The pipeline is designed for reductions of unresolved point sources.  However, for spectra of extended objects, MASE can still be used to bias-subtract, flat field and wavelength calibrate the observations.  Future versions of MASE will include a telluric correction routine, which will require additional calibration observations, and scripting capabilities for reducing multiple nights of data.
 
Support for the design and construction of the Magellan Echellette Specrograph
was received from the Observatories of the Carnegie Institution of Washington,
the School of Science of the Massachusetts Institute of Technology, and the
National Science Foundation in the form of a collaborative Major Research
Instrument grant to Carnegie and MIT (AST--0215989).   J.X.P. is partially supported by an NSF CAREER grant (AST--0548180).   We thank the referee for their constructive comments that improved this manuscript.  J.J.B. acknowledges Jackie Faherty and Dagny Looper for their illuminating conversations and extensive vetting of MAGE data. We thank Ricardo Covvarubias for early testing of the MAGE quicklook tool.  Finally, thanks to the entire Magellan staff, in particular Mauricio Martinez and Hern\'{a}n Nu\~{n}ez.


\begin{thebibliography}{27}
\expandafter\ifx\csname natexlab\endcsname\relax\def\natexlab#1{#1}\fi

\bibitem[{{Bernstein} {et~al.}(2003){Bernstein}, {Shectman}, {Gunnels},
  {Mochnacki}, \& {Athey}}]{2003SPIE.4841.1694B}
{Bernstein}, R., {Shectman}, S.~A., {Gunnels}, S.~M., {Mochnacki}, S., \&
  {Athey}, A.~E. 2003, in Society of Photo-Optical Instrumentation Engineers
  (SPIE) Conference Series, Vol. 4841, Society of Photo-Optical Instrumentation
  Engineers (SPIE) Conference Series, ed. M.~{Iye} \& A.~F.~M. {Moorwood},
  1694--1704

\bibitem[{{Bochanski} {et~al.}(2007){Bochanski}, {West}, {Hawley}, \&
  {Covey}}]{2007AJ....133..531B}
{Bochanski}, J.~J., {West}, A.~A., {Hawley}, S.~L., \& {Covey}, K.~R. 2007,
  \aj, 133, 531

\bibitem[{{Bolton} \& {Burles}(2007)}]{2007NJPh....9..443B}
{Bolton}, A.~S., \& {Burles}, S. 2007, New Journal of Physics, 9, 443

\bibitem[{{Cui} {et~al.}(2008){Cui}, {Ye}, \& {Bai}}]{2008AcASn..49..327C}
{Cui}, B., {Ye}, Z.~F., \& {Bai}, Z.~R. 2008, Acta Astronomica Sinica, 49, 327

\bibitem[{{Cushing} {et~al.}(2004){Cushing}, {Vacca}, \&
  {Rayner}}]{2004PASP..116..362C}
{Cushing}, M.~C., {Vacca}, W.~D., \& {Rayner}, J.~T. 2004, \pasp, 116, 362

\bibitem[{{Goodrich} \& {Veilleux}(1988)}]{1988PASP..100.1572G}
{Goodrich}, R.~W., \& {Veilleux}, S. 1988, \pasp, 100, 1572

\bibitem[{{Hall} {et~al.}(1994){Hall}, {Fulton}, {Huenemoerder}, {Welty}, \&
  {Neff}}]{1994PASP..106..315H}
{Hall}, J.~C., {Fulton}, E.~E., {Huenemoerder}, D.~P., {Welty}, A.~D., \&
  {Neff}, J.~E. 1994, \pasp, 106, 315

\bibitem[{{Hamuy} {et~al.}(1994){Hamuy}, {Suntzeff}, {Heathcote}, {Walker},
  {Gigoux}, \& {Phillips}}]{1994PASP..106..566H}
{Hamuy}, M., {Suntzeff}, N.~B., {Heathcote}, S.~R., {Walker}, A.~R., {Gigoux},
  P., \& {Phillips}, M.~M. 1994, \pasp, 106, 566

\bibitem[{{Hamuy} {et~al.}(1992){Hamuy}, {Walker}, {Suntzeff}, {Gigoux},
  {Heathcote}, \& {Phillips}}]{1992PASP..104..533H}
{Hamuy}, M., {Walker}, A.~R., {Suntzeff}, N.~B., {Gigoux}, P., {Heathcote},
  S.~R., \& {Phillips}, M.~M. 1992, \pasp, 104, 533

\bibitem[{{Horne}(1986)}]{1986PASP...98..609H}
{Horne}, K. 1986, \pasp, 98, 609

\bibitem[{{Kelson}(2003)}]{2003PASP..115..688K}
{Kelson}, D.~D. 2003, \pasp, 115, 688

\bibitem[{{Marsh}(1989)}]{1989PASP..101.1032M}
{Marsh}, T.~R. 1989, \pasp, 101, 1032

\bibitem[{{Marshall} {et~al.}(2008)}]{2008SPIE.7014E.169M}
{Marshall}, J.~L., {et~al.} 2008, in Society of Photo-Optical Instrumentation
  Engineers (SPIE) Conference Series, Vol. 7014, Society of Photo-Optical
  Instrumentation Engineers (SPIE) Conference Series

\bibitem[{{Massey} {et~al.}(1988){Massey}, {Strobel}, {Barnes}, \&
  {Anderson}}]{1988ApJ...328..315M}
{Massey}, P., {Strobel}, K., {Barnes}, J.~V., \& {Anderson}, E. 1988, \apj,
  328, 315

\bibitem[{{Moreno} {et~al.}(1982){Moreno}, {Llorente de Andres}, \&
  {Jimenez}}]{1982A&A...111..260M}
{Moreno}, V., {Llorente de Andres}, F., \& {Jimenez}, J. 1982, \aap, 111, 260

\bibitem[{{Mukai}(1990)}]{1990PASP..102..183M}
{Mukai}, K. 1990, \pasp, 102, 183

\bibitem[{{Murphy} {et~al.}(2007){Murphy}, {Tzanavaris}, {Webb}, \&
  {Lovis}}]{2007MNRAS.378..221M}
{Murphy}, M.~T., {Tzanavaris}, P., {Webb}, J.~K., \& {Lovis}, C. 2007, \mnras,
  378, 221

\bibitem[{{Oke}(1990)}]{1990AJ.....99.1621O}
{Oke}, J.~B. 1990, \aj, 99, 1621

\bibitem[{{Piskunov} \& {Valenti}(2002)}]{2002A&A...385.1095P}
{Piskunov}, N.~E., \& {Valenti}, J.~A. 2002, \aap, 385, 1095

\bibitem[{{Ponz} {et~al.}(1986){Ponz}, {Brinks}, \&
  {Dodorico}}]{1986SPIE..627..707P}
{Ponz}, D., {Brinks}, E., \& {Dodorico}, S. 1986, in Society of Photo-Optical
  Instrumentation Engineers (SPIE) Conference Series, Vol. 627, Society of
  Photo-Optical Instrumentation Engineers (SPIE) Conference Series, ed. D.~L.
  {Crawford}, 707--714

\bibitem[{{Robertson}(1986)}]{1986PASP...98.1220R}
{Robertson}, J.~G. 1986, \pasp, 98, 1220

\bibitem[{{Rossi} {et~al.}(1985){Rossi}, {Lombardi}, {Gaudenzi}, \& {de
  Biase}}]{1985A&A...143...13R}
{Rossi}, C., {Lombardi}, R., {Gaudenzi}, S., \& {de Biase}, G.~A. 1985, \aap,
  143, 13

\bibitem[{{Sheinis} {et~al.}(2002){Sheinis}, {Bolte}, {Epps}, {Kibrick},
  {Miller}, {Radovan}, {Bigelow}, \& {Sutin}}]{2002PASP..114..851S}
{Sheinis}, A.~I., {Bolte}, M., {Epps}, H.~W., {Kibrick}, R.~I., {Miller},
  J.~S., {Radovan}, M.~V., {Bigelow}, B.~C., \& {Sutin}, B.~M. 2002, \pasp,
  114, 851

\bibitem[{{Steinmetz} {et~al.}(2006)}]{2006AJ....132.1645S}
{Steinmetz}, M., {et~al.} 2006, \aj, 132, 1645

\bibitem[{{Stoughton} {et~al.}(2002)}]{2002AJ....123..485S}
{Stoughton}, C., {et~al.} 2002, \aj, 123, 485

\bibitem[{{SubbaRao} {et~al.}(2002){SubbaRao}, {Frieman}, {Bernardi},
  {Loveday}, {Nichol}, {Castander}, \& {Meiksin}}]{2002SPIE.4847..452S}
{SubbaRao}, M., {Frieman}, J., {Bernardi}, M., {Loveday}, J., {Nichol}, B.,
  {Castander}, F., \& {Meiksin}, A. 2002, in Society of Photo-Optical
  Instrumentation Engineers (SPIE) Conference Series, Vol. 4847, Society of
  Photo-Optical Instrumentation Engineers (SPIE) Conference Series, ed. J.-L.
  {Starck} \& F.~D. {Murtagh}, 452--460

\bibitem[{{York} {et~al.}(2000)}]{2000AJ....120.1579Y}
{York}, D.~G., {et~al.} 2000, \aj, 120, 1579

\end{thebibliography}
\end{document}